# Biphoton Spectrum Control

K. G. Katamadze and S. P. Kulik
*Moscow State University, Moscow, 119992 Russia*
*e-mail: katamadze@inbox.ru, Sergei.Kulik@gmail.com*

**Abstract** — The main methods for controlling the biphoton field, as well as the problems for which the width and the shape of the spectrum of the biphoton field are of decisive importance, are discussed. The method for controlling the spectrum of the spontaneous parametric downconversion of light based on the spatial modulation of the refractive indices of a nonlinear crystal in which the generation of biphotons has been analyzed. Modulation is due to the thermo-optic and electro-optic effects.

## 1. INTRODUCTION

One of the main aims of experimental quantum optics and quantum communication is the preparation of light in a given quantum state, when the parameters of the state are known a priori and/or can be controlled in the experiment.

The state of the biphoton field is specified by the spatial, spectral, and polarization parameters. In this work, we analyze methods for preparing biphoton fields with various frequency properties. Taking into account the spectral expansion, the state of two-photon light has the form [1, 2]

$$|\Psi\rangle = |vac\rangle + \zeta \iint d\omega_s d\omega_i F(\omega_s, \omega_i) a_s^\dagger(\omega_s) a_i^\dagger(\omega_i) |vac\rangle, \tag{1}$$

where $\omega_s$ and $\omega_i$ are the frequencies of the signal and idle photons, respectively; and $a_s^\dagger$ and $a_i^\dagger$ are the creation operators of photons in fixed signal and idler spatially polarized modes, respectively. Such a field is usually obtained using spontaneous parametric downconversion [3]. In the case of a narrowband pump[1], the frequencies $\omega_s$ and $\omega_i$ are related as $\omega_s + \omega_i = \omega_p$, where $\omega_p$ is the pump frequency. Then, it is convenient to represent the frequencies in the form of mismatches

$$\omega_s = \omega_{s0} + \Omega, \quad \omega_i = \omega_{i0} - \Omega, \tag{2}$$

where $\omega_{s0}$ and are the central frequencies of the spectra of the signal and idle fields ($\omega_{s0} + \omega_{i0} = \omega_p$), and represent Eq. (1) in the form

$$|\Psi\rangle = |vac\rangle + \zeta \int d\Omega F(\Omega) a_s^\dagger(\omega_{s0} + \Omega) a_i^\dagger(\omega_{i0} - \Omega) |vac\rangle. \tag{3}$$

Here, the generally complex function $F(\Omega)$, which is usually called the spectral amplitude of the biphoton, describes the frequency spectrum of the biphoton field.

Below, the spectrum of the biphoton field is primarily treated as its spectral amplitude and the width of the distribution of the function $|F(\Omega)|^2$ serves as the width of the spectrum, $\Delta\Omega$. Note that the spectral width of the biphoton in the case of collinear degenerate type-I phase matching is defined in [5, 6] as the length of the localization region of the amplitude $F(\omega_s, \omega_i)$ along the direction $\omega_s - \omega_i = 0$, which differs from the spectral width of the signal and idle photons $\Delta\Omega$ only by a factor $\sqrt{2}$.

The paper consists of two parts. The first part (Sections 2–4) reviews the main aims and methods for controlling the spectrum of the biphoton field. The second part (Sections 5–7) describes in detail the method for controlling the spectrum by means of the spatial modulation of the refractive indices of the nonlinear crystal. The new experimental results are also presented in the second part, where the method under investigation is compared with the methods reviewed in the first part.

---

[1] Hereinafter, we consider the case of the Fourier-limited pump for which the spectral width $\Delta\omega_p$ and pulse duration $\Delta\tau$ are related as $\Delta\omega_p \sim 1/\Delta\tau$. In this case, the pump can be treated as narrowband (i.e., can be approximated by the delta function) under the condition $2(v_g^{(p)} - v_g^{(s,i)})\Delta\tau / L > 1$ [4], where $v_g^{(p)}$ and $v_g^{(s,i)}$ are the group velocities

## 2. RELATION OF THE SPECTRUM OF THE BIPHOTON FIELD WITH THE CORRELATION CHARACTERISTICS

In most cases where the biphoton field is used, its correlation characteristics are of primary importance. For this reason, the control of the spectrum of the biphoton field is interesting in the context of the first and second order correlation functions. Note that the function $F(\Omega)$ is not measured directly in an experiment. However, the spectral intensities of the field in the signal and idler modes can be measured:

$$S_s \propto \left|F\left(\Omega = \omega - \omega_{s0}\right)\right|^2, \quad S_i \propto \left|F\left(\Omega = \omega - \omega_{i0}\right)\right|^2. \tag{4}$$

According to the Wiener–Khinchin theorem, the first-order correlation function for the signal and/or idle modes (i.e., for the single-photon field) is given by the expression

$$G^{(1)}(\tau) \propto \int |F(\Omega)|^2 \cos(\Omega\tau)d\Omega. \tag{5}$$

Under the conditions

$$F(-\Omega) = F(\Omega), \quad \omega_{s0} = \omega_{i0} = \omega_p/2 \tag{6}$$

the total spectral intensity of the field, $S = S_s + S_i$ has the form

$$S(\omega) \propto \left|F(\Omega = \omega - \omega_p/2)\right|^2, \tag{7}$$

In this case, the correlation function of the biphoton field is expressed similar to the single-photon correlation function given by Eq. (5) [1]. We emphasize that the width of the spectrum $S(\omega)$, which is related only to the absolute value of the amplitude $|F(\Omega)|$, completely determines the width of the correlation function $G^{(1)}(\tau)$:

$$\Delta^{(1)}\tau \sim 1/\Delta\omega = 1/\Delta\Omega \tag{8}$$

The state of the biphoton can be described not only by the spectral amplitude, but also by the time amplitude

$$\tilde{F}(\tau) = \int d\Omega e^{i\Omega\tau} F(\Omega), \tag{9}$$

whose absolute value squared provides the second-order correlation function [1]:

$$G^{(2)}(\tau) \propto \left|\int F(\Omega)\cos(\Omega\tau)d\Omega\right|^2, \tag{10}$$

This expression does not require condition (6). An important difference of Eq. (10) from Eq. (5) is that the width of the second-order correlation function $\Delta^{(2)}\tau$ is determined not only by the absolute value, but also the phase of the amplitude of $F(\Omega)$. The relation $\Delta^{(2)}_{min}\tau \sim 1/\Delta\omega = 1/\Delta\Omega$ is satisfied only for the minimum possible value of the width of $\Delta^{(2)}\tau$, which is reached when the phase of $F(\Omega)$ slightly depends on the frequency. In this case, the broad spectrum of the biphoton field is a necessary, but not sufficient condition for the smallness of the time $\Delta^{(2)}\tau$. At the same time, the biphoton field with a narrow spectrum always has a wide second-order correlation function.

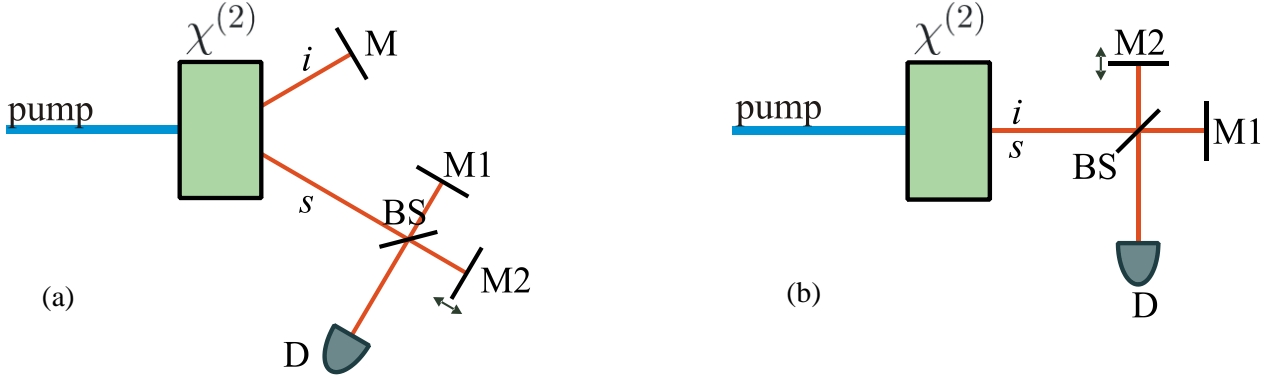

**Fig. 1. Scheme of the measurement of the first-order correlation function using a Michelson interferometer.**
Radiation generated in a crystal with a nonlinear susceptibility $\chi^{(2)}$ is guided to the interferometer consisting of beam splitter BS and two mirrors M1 and M2; after that, the radiation intensity is measured by detector D. Interference can be observed by displacing mirror M2. The width of the function $G^{(1)}(\tau)$ can be estimated from a change in the visibility of the interference pattern.
(a) Measurement of $G_s^{(1)}(\tau)$ in signal mode; (b) Measurement of $G^{(1)}(\tau)$ in both modes.

The correlation functions $G^{(1)}(\tau)$ and $G^{(2)}(\tau)$, as well as the spectral intensity, can also be measured in an experiment. The first-order correlation function is manifested in interference periments. For example, if the signal and idle photons are in different spatial modes, the correlation function $G^{(1)}(\tau)$ can be measured with a Michelson interferometer placed in either signal or idle mode (see Fig. 1a). In the case of the collinear degenerate regime and the same polarization states of photons, the correlation function $G^{(1)}(\tau)$ of the biphoton field can be measured with the same interferometer (see Fig. 1b) [7].

One of the remarkable effects that are consequences of the quantum properties of the biphoton field is the so-called Hong–Ou–Mandel dip: if both photons of a pair arrive simultaneously on both inputs of the 50% beam splitter, they come to the same arm in the output if they are absolutely indistinguishable. This indistinguishability includes shability in time. Spontaneous parametric downconversion is accompanied by the generation of photons correlated in the creation time and, for they to remain correlated to the time of their arrival at the beam splitter, the lengths of their optical paths should coincide with each other with an accuracy of the inverse spectral width[2]. The effect is experimentally observed in coincidences of photocounts of the detectors placed in the output modes of the beam splitter (see Fig. 2). The length of one of the optical paths is varied and, when the lengths of both paths are the same, a dip appears in dences because both photons are always guided

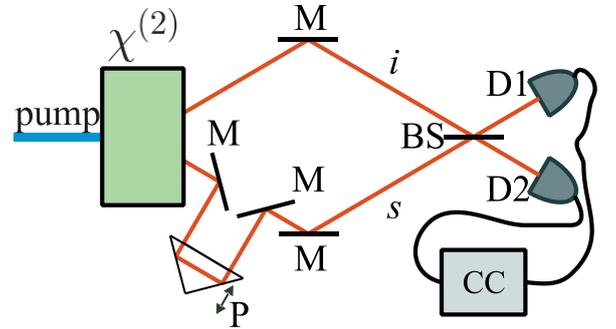

**Fig. 2. Scheme of the measurement of the first-order correlation function by measuring the Hong–Ou–Mandel dip.**
The biphoton radiation is generated in the crystal with the nonlinear susceptibility $\chi^{(2)}$ in the noncollinear regime degenerate in the frequency and polarization. Using mirrors M, the signal (s) and idle (i) modes are mixed on a 50% beam splitter BS. Then, radiation is guided to detectors D1 and D2, which are connected through the coincidence scheme CC. The length of the signal channel can be changed by displacing prism P. When the optical paths of the signal and idle photons are the same, they are shared with a probability of 100% to one of two outputs of the beam splitter and the coinciding photocounts will exhibit a dip whose width corresponds to the half-width of the correlation function $G^{(1)}(\tau)$ given by Eq. (12).

---

[2] Note that there are a number works in the field of quantum optics concerning the so-called postponed compensation in which the manifestation of two-photon interference in the Hong–Ou–Mandel scheme is not associated with the simultaneous arrival of two photons at the beam splitter [8].

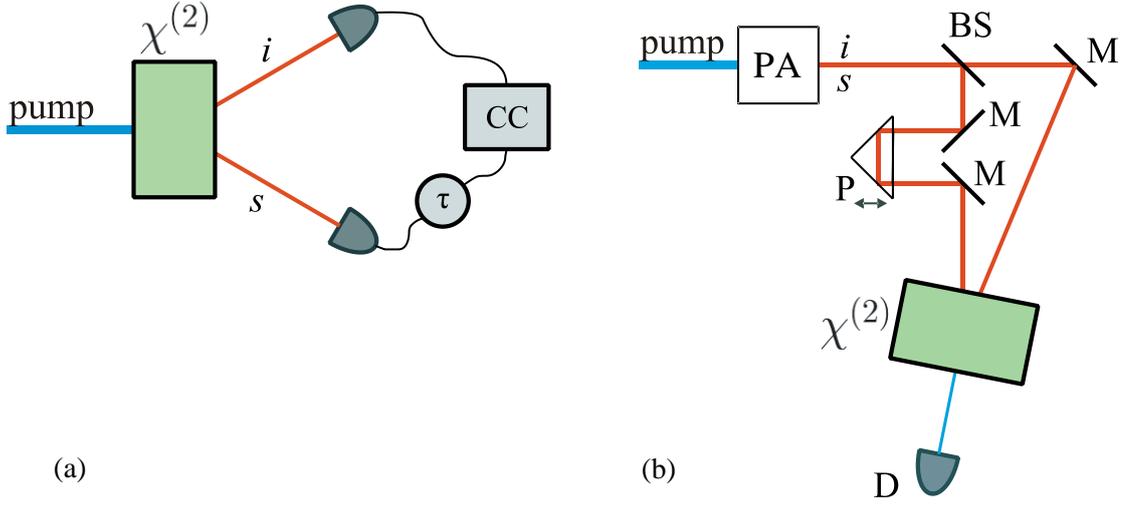

**Fig. 3. Scheme of the measurement of the second-order correlation function.**

(a) Scheme of the measurement of the correlation function $G^{(2)}(\tau)$ in a Hanbury Brown–Twiss interferometer. The signal *s* and idler *i* photons obtained in a nonlinear crystal in the presence of spontaneous parametric downconversion are guided in different spatial modes and are detected by the detectors. The function $G^{(2)}(\tau)$ can be measured by varying the delay $\tau$ of signals propagating from of the detectors to coincidence scheme CC (if the window of the coincidence scheme is much narrower than the width $\Delta^{(2)}(\tau)$).

(b) Scheme of the measurement of the correlation function $G^{(2)}(\tau)$ due to the generation of the sum frequency. Radiation from parametric amplifier PA is separated into two channels by beam splitter BS; after that, mirror M guides both beam into a nonlinear crystal in which the second-harmonic generation occurs in the noncollinear regime. Radiation at doubled frequency is registered by detector D. The second-order correlation function can be measured by varying the length of one of the optical paths through the displacement of prism P. In this case, the nonlinear crystal serves as a precision coincidence scheme.

to one detector. It is known that, if the window of the coincidence scheme is sufficiently wide, the shape and width of this dip are related to the first-order correlation function as [9, 10]

$$R_c^{(2)}\tau \sim 1 - g^{(1)}(2\tau), \tag{11}$$

where the function $R_c(\tau)$ describes the dependence of the coincidence count rate on the time delay introduced for one of the photons and $g^{(1)}(\tau)$ is the normalized first-order correlation function for a pair of photons.

The second-order correlation function is manifested in correlation between the counts in the detectors detected signal and idle photons. When the window of the coincidence scheme is much narrower than $\Delta^{(2)}\tau$, the function $G^{(2)}(\tau)$ can be measured with a Hanbury Brown–Twiss interferometer (see Fig. 3a) by varying the time delay $\tau$ between pulses arriving the coincidence scheme and measuring the count rate of the coincidences of photocounts $R_c(\tau) \sim G^{(2)}(\tau)$. However, $\Delta^{(2)}\tau$ is in most cases much smaller than the window of the coincidence scheme; in these cases, such measurements are impossible. The function $G^{(2)}(\tau)$ is also manifested in two-photon interactions with matter, as well as in parametric processes, e.g., in the generation of sum-frequency radiation. The probability of such process is proportional to the second-order correlation function. The authors of [11] considered an experiment (see Fig. 3b) in which light from a parametric amplifier operating in the collinear degenerate regime was split by a beam splitter into two channels; after that, a controlled delay was introduced to one of the channels and the beams were reunited on a nonlinear crystal in which the second-harmonic generation occurred in the noncollinear regime. The intensity of the double-frequency radiation was measured as a function of the delay, $I_{2\omega}(\tau) \sim G^{(2)}(\tau)$.

## 3. PROBLEMS IN WHICH THE INCLUSION OF THE SPECTRUM OF BIPHOTON FIELD IS IMPORTANT

In a number of applications, a source of the biphoton field with a narrow spectrum is required. First, to increase the efficiency of the single-photon interactions of light with individual atoms and to implement quantum memory [12, 13], the frequency of photons should be in resonance with energy levels. Since one of the basic elements of the scheme (heralded scheme) for generating single-photon states is the biphoton field [14–17], its typical spectral width for these problems should be no more than 1–10 MHz.[3]. Second, in the case of the transmission of quantum information through optical fibers, the arrival time of single-photon packages is smeared in view of chromatic dispersion; for this reason, the spectral composition of the packages should be minimized [18]. In addition, the biphoton field with a narrow spectrum (and, therefore, with a large correlation times necessary for measuring the time characteristics of single-photon detectors [19], which can be determined in the Hanbury Brown–Twiss scheme (see Fig. 3a).

The biphoton field with a broad spectrum is used in another group of applications. Here, we primarily point to the problem of increasing the entanglement degree of the biphoton field, which is of current importance for quantum information encoding. An increase in the entanglement degree is accompanied by an increase in the effective dimension of the Hilbert space of strongly correlated optical states, which is a promising object of quantum cryptography [20–23] and for testing the fundamentals of quantum theory [24–27].

When the biphoton field is in a pure state of form (3), the parameter $R$, that was introduced by M.V. Fedorov [4, 28] and is defined as the ratio of the spectral width of individual photons (unconditional distribution) to the spectral width of coincidences of photocounts (conditional distribution) [29] is very convenient for the numerical analysis of the entanglement degree. Since the spectral width of coincidences in spontaneous parametric scattering for the narrowband pump is determined by the spectral width of the pump, the spectral width of single counts coincides with the width of the function $|F(\Omega)|^2$. Thus, the frequency entanglement degree under a given pump can be increased only by the broadening of the spectrum of the biphoton field.

As was mentioned above, the broadening of the spectrum of the biphoton field leads to a decrease in the correlation time $\Delta^{(1)}_{s,i}\tau$ between photons in each mode; this fact can be used in optical coherent tomography (see Fig. 4a). In optical coherent tomography, an interferometer with an object under investigation in one of the arms is placed in one of the modes of the biphoton field. The object can be scanned at various depths by varying the length of the second arm of the interferometer. The resolution of optical coherent tomography is determined by the radiation coherence length $l_{res} = c / \Delta^{(1)}_{s,i} \tau$. An alternative of optical coherent tomography is the quantum optical coherent tomography [30] based on the Mandel dip. The object under investigation is placed in one of the channels in front of the beam splitter (see Fig. 4b) and the optical length of the other channel is varied. According to Eq. (11), the resolution of such a scheme is determined by double coherence time, $l_{res} = c / 2\Delta^{(1)} \tau$. Note that quantum optical coherent tomography requires a source of the biphoton field that satisfies conditions (6) and whose signal and idle modes differ in the spatial or polarization parameters.

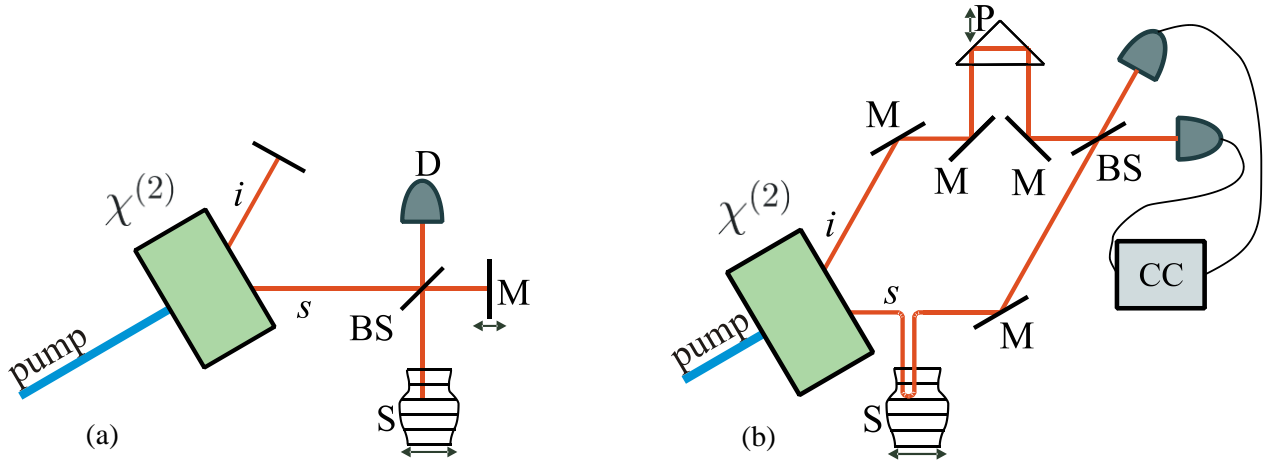

**Fig. 4. Optical tomography types.**
(a) Optical coherent tomography. Radiation was fed to a Michelson interferometer consisting of beam splitter BS and mirror M. Sample S was placed in one of the arms of the interferometer. Light at the output of the interferometer is detected by detector D. The sample can be scanned at different depths by varying the position of the mirror.
(b) Quantum optical coherent tomography. A pair of photons is created in a nonlinear crystal. Sample S is placed in signal mode *s* and the length of the path of idle photon *i* is varied by prism P. Taking photons on beam splitter BS, images of different layers of the sample can be obtained using the Hong–Ou–Mandel dip effect. The scanning depth can be varied by displacing the prism.

In addition, the biphoton field with a broad spectrum is necessary for problems requiring the efficient biphoton interaction of light with matter. In the case of short times $\Delta^{(2)} \tau$, the biphoton behaves as a united object with the effective wavelength [31]

$$\lambda_{eff} = \frac{2\pi c}{E/\hbar} = \frac{2\pi c}{2\hbar\omega/\hbar} = \frac{1}{2}\frac{2\pi c}{\omega} = \frac{1}{2}\lambda, \qquad (12)$$

where $E$ — is the energy of the biphoton and $\omega$ is the angular frequency of constituent photons (in the frequency-degenerate case). This fact is used to increase the resolution of nonlinear microscopy [32] and quantum interference optical lithography [33]. The spectroscopy of virtual states with the use of entangled photons is based on two-photon absorption [34].

Note the problem of the synchronization of clocks for solving which the use of a pair of photons with small coherence time $\Delta^{(2)} \tau$ was proposed [35] when a pair of photons is separated into two spatial modes; then, the photons were guided to detectors triggering clocks that should be synchronized.

We emphasize that most applications discussed above impose requirements only on the spectral width of the biphoton field, not on its shape. It is always assumed that the spectral shape is close to Gaussian or rectangular; however, this is not necessarily the fact in reality. In view of this circumstance, the problem of controlling not only the width, but also the shape of the spectrum of biphotons is interesting.

## 4. METHODS FOR CONTROLLING THE SPECTRUM OF THE BIPHOTON FIELD

Two-photon light is usually generated owing to spontaneous parametric scattering. As was mentioned above, the state of the field in the case of a narrowband pump at the output of the nonlinear crystal can be represented in the form of Eq. (3), where the spectral amplitude $F(\Omega)$ is given by the expression [2]

$$F(\Omega) \propto \int_0^L dz \exp[i\Delta k(\Omega)z], \qquad (13)$$

where $z$ is the coordinate along the axis directed along the pump propagation, $L$ is the length of the crystal in which generation occurs, and $\Delta \vec{k} = \vec{k}_p - \vec{k}_s - \vec{k}_i$ is the. phase detuning. In the case of the homogeneous crystal, is independent of $z$ and Eq. (13) is simplified as

$$F(\Omega) \propto L \exp\left(-i\frac{\Delta k(\Omega)L}{2}\right) \text{sinc}\left(\frac{\Delta k(\Omega)L}{2}\right). \qquad (14)$$

According to Eq. (14), it is seen that the spectrum of the biphoton field is determined by the frequency dependence of the phase detuning $\Delta k(\Omega)$ and its width, by the condition

$$-\frac{2\pi}{L} \leq \Delta k(\Omega) \leq \frac{2\pi}{L}. \qquad (15)$$

Thus, in order to obtain the biphoton field with a narrow spectrum, a sufficiently long crystal can be taken and be placed into the cavity for additional frequency selection. In this case, the spectrum narrower than 3.0 MHz can be obtained [36].

The problem of obtaining the biphoton field with a broad spectrum is more difficult. As follows from Eq. (15), the trivial solution would be the use of a short crystal. In particular, a field with a spectral width of 174 nm (106 THz[4]) was obtained in [37] with the use of the BBO crystal 0.1 mm in thickness in the orthogonal polarization modes in the degenerate regime at a wavelength of 703 nm. However, the intensity of radiation proportional to the square of the length of the crystal, Eq. (14), decreases in the case of the thin crystal. The integral intensity decreases linearly with a decrease in $L$ [38].

Another method for obtaining the broad spectrum of the biphoton field is the choice of the matching conditions such that the detuning function $\Delta k(\Omega)$ depends weakly on $\Omega$ in a certain interval near the exact matching ($\Delta k = 0$). By definition, $\Delta k(\Omega) = k_p - k_s - k_i$ or, in representation (2),

$$\Delta k(\Omega) = k_p - k_s(\omega_{s0} + \Omega) - k_i(\omega_{i0} - \Omega) = k_p - k_{s0}(\Omega) - k_{i0}(-\Omega), \tag{16}$$

where

$$k_{s,i0}(\Omega) = k_{s,i}(\omega_{s,i0} + \Omega). \tag{17}$$

Expanding $\Delta k(\Omega)$ in a Taylor series, we obtain

$$\Delta k(\Omega) = \left[k_p - k_{s0} - k_{i0}\right] - \left[k'_{s0} - k'_{i0}\right]\Omega - \frac{1}{2}\left[k''_{s0} + k''_{i0}\right]\Omega^2 - \ldots, \tag{18}$$

where all derivatives of the functions $k_{s,i0}(\Omega)$ are taken at zero.

Consequently, the broadband matching requires the conditions

$$k_p - k_{s0} - k_{i0} = 0, \tag{19}$$

$$k'_{s0} - k'_{i0} = 0, \tag{20}$$

$$k''_{s0} + k''_{i0} = 0. \tag{21}$$

The first condition is the exact phase matching condition for the central frequencies of the signal and idle photons, the second condition means the equality of their group velocities, and the third condition ensures the absence of the dispersion of group velocities. Note that conditions (19) and (20) are satisfied automatically for the degenerate matching of type I, because the functions $k_{s0}(\Omega)$ and $k_{i0}(\Omega)$ are identical in this case and $\Delta k_I \propto \Omega^2$. In the case of nondegenerate matching or type-II matching, where the polarizations of the signal and idle photons are orthogonal, $\Delta k_{II} \propto \Omega$. However, $\Delta k_I \propto \Omega^4$, for degenerate type-I matching under condition (21) and $\Delta k_{II} \propto \Omega^2$ for type-II matching under condition (20).

It is very difficult, though possible to choose the medium simultaneously satisfying conditions (19)–(21), i.e., to ensure the local weakening of the $\Delta k(\Omega)$ dependence near the exact satisfaction of the phase matching condition. In particular, it was shown in [39] that the width of collinear degenerate type-I matching for a 14-mm-thick BBO crystal and a 728-nm pump is about 750 nm (106 THz).

---

[4] As a characteristic of the spectral width of the biphoton field, the range width is taken, because it is certainly related to the correlation times $\Delta^{(1)}\tau$ and $\Delta^{(2)}_{min}\tau$, the short correlation time is responsible for the value of the broad spectrum in most applications.

To simplify the problem of simultaneous satisfaction of conditions (19)–(21), it was proposed to use periodically polarized crystals in which the phase matching condition is satisfied taking into account the reciprocal vector $\vec{k}_g$:

$$\Delta\vec{k} = \vec{k}_p - \vec{k}_s - \vec{k}_i - \vec{k}_g, \qquad (22)$$

where $k_g = 2\pi/\Lambda$, and $\Lambda$ is the period of the polarization of the square susceptibility of the crystal. Then, condition (19) is modified to the form

$$k_p - k_{s0} - k_{i0} - k_g = 0, \qquad (23)$$

which makes it possible to satisfy conditions (20) and (21) by choosing the dispersion of the crystal and the wavelength of the pump and to satisfy condition (23) by choosing the period of induced polarization. It was shown in [40] that the generation of the biphoton field with a spectral width of 1080 nm (91 THz) can be ensured in a periodically polarized lithium niobate (LiNbO$_3$) crystal 1 cm in thickness with a period of $\Lambda = 27.4$ μm under the condition of collinear type-I matching degenerate at a wavelength of 1885 nm.

Another method for locally weakening the $\Delta k(\Omega)$ dependence was demonstrated in [5]. The feature of the proposed scheme is the use of elements introducing angular dispersion (see Fig. 5). It was shown in [41–43] that a light pulse passing through a system of two diffraction gratings (or prisms) surrounding a medium with the walk-off effect is transformed as if it propagated through a medium with changed derivatives of the dispersion function, $k'(\omega)$ and $k''(\omega)$:

$$\tilde{k}' = k' + \alpha\rho, \qquad \tilde{k}'' = k'' - \frac{\alpha^2}{k}, \qquad \rho = tg\theta, \qquad \alpha = \frac{tg\varphi}{c}, \qquad (24)$$

where $\varphi$ dependence was demonstrated in [5]. The feature of the proposed scheme is the use of elements introducing angular dispersion (see Fig. 5). It was shown in [41–43] that a light pulse passing through a system of two diffraction gratings (or prisms) surrounding a medium with the walk-off effect is transformed as if it propagated through a medium with changed derivatives of the dispersion function, $\theta$ is the inclination angle of the pulse front [44], which appears after the first dispersing element and is compensated by the second dispersing element, and $c$ is the speed of light. The angle depends on the parameters of the dispersing element and on the central wavelength of the pulse. Using Eqs. (24), one can choose the dispersing elements such as to ensure conditions (20) and (21) [5]. In particular, the spectral width in a BBO crystal 2 mm in thickness was experimentally increases from 5.2 nm (2.4 THz) to 41 nm (19 THz) for the case of collinear type-II matching degenerate at a wavelength of 810 nm. The broadening of the spectrum from 96 nm (44 THz) to 465 nm (213 THz) for the case of degenerate type-I matching was theoretically predicted. However, the correctness of approximations (24) used for such a broad spectral range is doubtful in view of the initial limit of the spectral width.

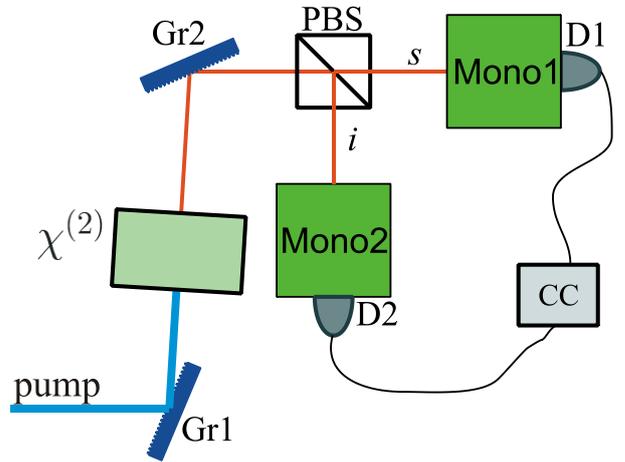

**Fig. 5. Local smoothing of the $\Delta k(\Omega)$ dependence due to angular dispersion.**
The generation of spontaneous parametric downconversion in type-II phase matching occurs in a nonlinear crystal located between two diffraction gratings. Grating Gr1 creates an inclination of the pump pulse front, which holds for the pulse front of spontaneous parametric downconversion and is compensated by lattice Gr2. As a result, the phase matching conditions are modified to Eqs. (25). Polarization beam splitter PBS separates the signal and idle photons into two spatial modes, which are then registered by detectors D1 and D2. To determine the spectrum of the biphoton field, monochromators Mono1 and Mono2 are used.

Additional possibilities of varying the function $\Delta k(\Omega)$ appear in the case of non-collinear matching. Let us consider the generation of spontaneous parametric scattering in a periodically polarized crystal by a monochromatic pump having a certain angular distribution $f(q)$ with a width of $\Delta q$, where $q$ is the transverse component of the wave vector (see Fig. 6) [45]. It is convenient to represent the phase detuning $\Delta \vec{k}$ in the form of the sum of two components $\Delta \vec{k}_\perp$ and $\Delta \vec{k}_\parallel$ perpendicular and parallel to the $z$ axis, respectively, coinciding with the pump propagation direction. In this case, the phase matching conditions have the form:

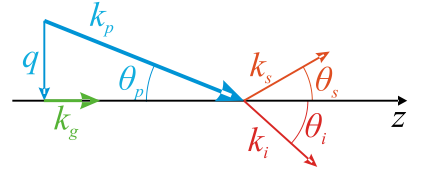

Fig. 6. Orientation of the wave vectors under the non-collinear phase matching of spontaneous parametric downconversion in a periodically polarized sample.

$$\Delta \vec{k}_\perp = 0, \quad -\frac{2\pi}{L} \leq \Delta k_\parallel \leq \frac{2\pi}{L}. \tag{25}$$

In the approximation of the narrow angular spectrum of the pump $q \equiv k_{p\perp} = k_p \sin\theta_p \approx k_p \theta_p$, $k_{p\parallel} = k_p \cos\theta_p \approx k_p$. In this case, $\Delta \vec{k}_\perp$ and $\Delta \vec{k}_\parallel$ can be represented in the form

$$\Delta k_\perp = q + k_s \sin\theta_s - k_i \sin\theta_i, \qquad \Delta k_\parallel = k_p - k_s \cos\theta_s - k_i \cos\theta_i - k_g. \tag{26}$$

Taking into account representation (16), we expand Eqs. (26) into the Taylor series in $\Omega$, in the form similar to Eq. (18):

$$\begin{aligned}\Delta k_\perp(\Omega) = q &+ [k_{s0}\sin\theta_s - k_{i0}\sin\theta_i] + \\ &+ [k'_{s0}\sin\theta_s + k'_{i0}\sin\theta_i]\Omega + \\ &+ 1/2 [k''_{s0}\sin\theta_s - k''_{i0}\sin\theta_i]\Omega^2 + ...\end{aligned} \tag{27}$$

$$\begin{aligned}\Delta k_\parallel(\Omega) = &[k_p - k_g - k_{s0}\cos\theta_s - k_{i0}\cos\theta_i] - \\ &- [k'_{s0}\cos\theta_s - k'_{i0}\cos\theta_i]\Omega - \\ &- 1/2\ [k''_{s0}\cos\theta_s + k''_{i0}\cos\theta_i]\Omega^2 + ...\end{aligned} \tag{28}$$

According to Eq. (28), the $\Delta k_\parallel(\Omega)$ is weakened with an increase in the angles $\theta_s$ and $\theta_i$. At the same time, the condition $\Delta \vec{k}_\perp = 0$ together with Eq. (27) implies the constraint on the spectral width, which is attributed to the width of the angular distribution of the pump $\Delta q$, and is enhanced with an increase in the angles $\theta_s$ and $\theta_i$. Let us consider the case of degenerate type-I matching, where the signal and idle modes have the ordinary polarization. In this case, $k_{s0} = k_{i0} = k_0$, $\theta_s = \theta_i = \theta_0$, and under the exact matching condition

$$k_p - k_g - k_{s0}\cos\theta_s - k_{i0}\cos\theta_i = 0, \tag{29}$$

Eqs. (27) and (29) have the form

$$\Delta k_\perp = q + 2k'_0 \sin\theta_0 \Omega + ... \tag{30}$$

$$\Delta k_\parallel = -2k''_0 \cos\theta_0 \Omega^2 + ... \tag{31}$$

The condition $\Delta k_\perp = 0$ implies the constraint on the spectral width:

$$\Omega \leq \frac{1}{2k'_0 \sin\theta_0}\Delta q, \tag{32}$$

and the condition $\Delta k_\parallel \leq \frac{2\pi}{L}$ provides:

$$\Omega \leq \sqrt{\frac{2\pi}{L k_0'' \cos \theta_0}}. \tag{33}$$

Thus, the spectrum can be strongly broadened for the strongly focused pump radiation. In particular, a pump-focusing-induced increase in the spectral width from 6.2 nm (2.8 THz) to 148 nm (67 THz) was experimentally demonstrated in [46] for the case of degenerate type-I matching at a wavelength of 812 nm. Note that this method leads to angular broadening, which is a direct consequence of the matching conditions.

Note that the matching conditions for the case of the transverse matching ($\theta_s = \theta_i = 90°$) are degenerate and have the form

$$\Omega = \frac{1}{2k_0'} q, \tag{34}$$

i.e., the form of the frequency spectrum of the biphoton field under closed matching condition completely corresponds to the angular spectrum of the pump. This provides the possibility of controlling the shape of the frequency spectrum.

The spectrum width of the biphoton field can be increased due to the broadening of not only the angular, but also frequency spectrum of the pump. In this case, the frequency anticorrelation condition is softened ($\omega_s + \omega_i \approx const$), but the small broadening of the frequency spectrum of the pump under certain conditions can lead to the strong broadening of the spectrum of biphotons. Taking into account the width of the pump spectrum, Eqs. (2) are represented in the form

$$\omega_p = \omega_{p0} + \Omega_p, \quad \omega_s = \omega_{s0} + \Omega_s, \quad \omega_i = \omega_{i0} - \Omega_i, \tag{35}$$

where the frequency matching condition is satisfied both for the central frequencies, $\omega_{p0} = \omega_{s0} + \omega_{i0}$, and for detunings, $\Omega_p = \Omega_s - \Omega_i$. Let us expand $\Delta k$ in a power series in $\Omega_p$, $\Omega_s$, and $\Omega_i$. Hereinafter, we retain only the first power of $\Omega_p$, assuming that the spectral width of the pump is much smaller than the spectral width of biphotons:

$$\Delta k = \left[ k_{p0} - k_{s0} - k_{i0} \right]_0 + \left[ k_{p0}' \Omega_p - k_{s0}' \Omega_s + k_{i0}' \Omega_i \right]_1 - \frac{1}{2} \left[ k_{s0}'' \Omega_s^2 + k_{i0}'' \Omega_i^2 \right]_2 - \ldots \tag{36}$$

Let us consider the case of degenerate type-I matching ($k_{s0} = k_{i0} = k_0$). Taking into account that $\left[ k_{p0} - k_{s0} - k_{i0} \right]_0 = 0$ under the exact satisfaction of the phase matching condition for central frequencies, we obtain

$$\Delta k = \left( k_{p0}' - k_0' \right) \Omega_p - k_{s0}'' \left[ \Omega_s^2 - \Omega_p \Omega_s \right]_2. \tag{37}$$

The condition $\Delta k = 0$ provides the square equation with respect to $\Omega_s$, whose solution has the form [4, 6]

$$\Omega = \frac{\Omega_p}{2} \pm \sqrt{\gamma \Omega_p} \approx \sqrt{\gamma \Omega_p}, \text{ where } \gamma = \frac{k_{p0}' - k_0'}{k_0''}. \tag{38}$$

Note that the coefficient $\gamma$ is nonnegative for the case of normal dispersion. Thus, the width of the spectrum of the biphoton field is $\Delta \Omega = \sqrt{\gamma \Delta \Omega_p}$ for the fixed spectral width of the pump $\Delta \Omega_p$.

The broadening of the spectrum to 197 nm (84 THz) at a central wavelength of 840 nm was experimentally demonstrated in [47], where the spectral width of the pump was 7.7 THz and lithium iodate (LiIO$_3$) was used as a nonlinear crystal. A similar broadening of the spectrum in the nondegenerate regime at the wavelengths of 741 and 909 nm to about 30 THz, which is much larger than the width of the typical spectrum of spontaneous parametric scattering in the nonde-generate regime, was demonstrated in [48], where a BBO crystal was used as the nonlinear crystal 3 mm in length and the spectral width of the pump was 6.5 THz.

We emphasize that all considered methods for broadening the spectrum of the biphoton field are reduced to the local weakening of the $\Delta k(\Omega)$ dependence, which ensures the satisfaction of the matching conditions in a wider wavelength range. This leads, on one hand, to the broadening of the absolute value of the spectral amplitude and, on the other hand, to a weak frequency dependence of the spectral amplitude. This makes it possible to narrow not only the first-order correlation function, but also the second-order correlation function. Nevertheless, such a method for broadening of the spectrum of spontaneous parametric scattering is strongly restricted by the dispersion relations in the medium and is applicable only for the case of comparatively small frequency detunings $\Omega$, when the several first terms of the Taylor expansion are sufficient. Moreover, this method is inapplicable for the nondegenerate regime.

Recall that Eq. (14) for the spectral amplitude was obtained under the assumption that the crystal in which spontaneous parametric scattering occurs is spatially homogeneous and the phase detuning $\Delta k$ is independent of $z$. The use of spatially inhomogeneous structures makes it possible to simultaneously satisfy the matching conditions for different-frequency pairs in different regions of the crystal. As a result, spontaneous parametric scattering appearing in different parts of the crystal is added taking into account phases and output radiation has a broad spectrum with a complex shape (due to interference) and a nontrivial frequency dependence of the spectral amplitude:

$$F(\Omega) \propto \int_0^L dz \exp[i\Delta k(\Omega,z)z]. \tag{39}$$

Thus, the resulting radiation can be Fourier limited and, to reduce $\Delta^{(2)}\tau$ compression methods should be additionally used [49–51].

To obtain the spatial dependence $\Delta k(z)$ it is possible to use periodically polarized crystals, where polarization period increases [52] so as to ensure a linear increase (chirp) in the reciprocal vector (see Fig. 7a):

$$k_g(z) = k_{g0} + \alpha z. \tag{40}$$

In this case, the detuning has the form

$$\Delta k(\Omega,z) = k_p - k_{s0}(\Omega) - k_{i0}(-\Omega) - k_g(z). \tag{41}$$

The spectral width of the resulting radiation can be estimated as the difference

$$\tilde{\Omega}(z=0) - \tilde{\Omega}(z=L), \tag{42}$$

where $\tilde{\Omega}(z)$ is the solution of the equation $\Delta k(\Omega,z) = 0$. In particular, parametric radiation at a degenerate wavelength of 812 nm with a spectral width from 17 nm (7.7 THz) to 300 nm (136 THz) was obtained in [53] when the growth parameter α in the stoichiometric lithium tantalate 18 mm in length increases from $0.2\times10^{-7}$ to $9.7\times10^{-6}$ μm. In that work, the spectrum had a complex shape and included several peaks (see Fig. 7b).

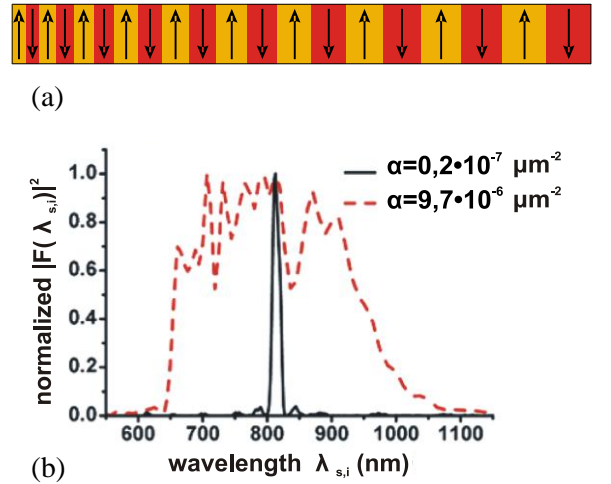

**Fig. 7.** Broadening of the spectrum in the periodically polarized structure with linear chirp [53].
(a) Periodically polarized crystal with a linear variation of the polarization period in the radiation propagation direction. The arrows indicate the directions of the polarization in different parts of the crystal.
(b) The wavelength dependence of the spectral intensity of the biphoton field at different values of $a$.

Radiation obtained in hyperparametric scattering [35] can be used as an alternative source of the biphoton field with a broad spectrum. Since hyperparametric scattering is due to the third-order nonlinear susceptibility $\chi^{(3)}$, it can be observed in media with an inversion center, e.g., in an optic fiber. The smallness of $\chi^{(3)}$ can be compensated by the length of the fiber. Such a method for obtaining the biphoton field has certain advantages as compared to the methods discussed above, because radiation obtained in the fiber can be matched with other fiber optical devices. The radiation of spontaneous parametric scattering obtained in the crystal is not diffraction limited and it is difficult to introduce it in the fiber with small losses. Moreover, matching conditions for hyperparametric scattering are much weaker and the matching width is much larger. The problem is that it is difficult to detect hyperparametric scattering in the degenerate regime, because emission occurs at the pump frequency. In addition, there are a number of incidental effects induced by $\chi^{(3)}$ nonlinearity that are comparable in intensity with hyperparametric scattering; Raman scattering is the strongest of these effects. Separating hyperparametric scattering radiation from "supercontinuum," one can obtain a spectrum with a width of up to 10 THz at a detuning of 25 THz from the degenerate regime at a wavelength of 741 nm [54].

## 5. INHOMOGENEOUS BROADENING OF THE SPECTRUM DUE TO THE SPATIAL MODULATION OF THE REFRACTIVE INDICES ALONG THE NONLINEAR CRYSTAL

Another method for creating the dependence of $\Delta k$ on $z$ was proposed in [52]. It consists of varying the refractive indices $n_p$, $n_s$ and $n_i$, using their dependence on the external parameters such as temperature [55], electrostatic field, and pressure. This method is considered experimentally and theoretically in this work.

1. Let us consider the control of the spectrum due to the temperature modulation of the refractive index. The temperature distribution $T(z)$, induced along the crystal leads to the spatial dependence $\Delta k(T(z))$; consequently, the integral amplitude is given by the integral

$$F(\Omega) \propto \int_0^L dz \, exp\left[i\Delta k\left(\Omega, T(z)\right)z\right]. \tag{43}$$

The temperature dependence of the phase detuning $\Delta k$ is described as follows. Let the refractive indices be linear functions of the temperature,

$$n_j = n_{j0} + \eta_j T, \quad \text{where} \quad j = p, s, i. \tag{44}$$

For simplicity, we assume that $\eta_s \approx \eta_i \equiv \eta$, which is valid near the degenerate regime. In this case, the wave detuning is given by the expression

$$\Delta k = \Delta k_0 + \frac{\omega_p}{c}(\eta_p - \eta)T. \tag{45}$$

If the temperature of the sample is a linear function of the longitudinal coordinate, $T(z) = T_0 + \gamma z$, the dependence $\Delta k(z)$ has the form

$$\Delta k(z) = \Delta k_0 + \delta z, \quad \text{where} \quad \delta = \frac{\omega_p}{c}\gamma. \tag{46}$$

Thus, the detuning depends not only on the parameters of the crystal and pump, but also on the external control parameter $\gamma$.

To estimate the spectral width, modified expression (42) can be used:

$$\tilde{\Omega}(T(z=0)) - \tilde{\Omega}(T(z=L)), \tag{47}$$

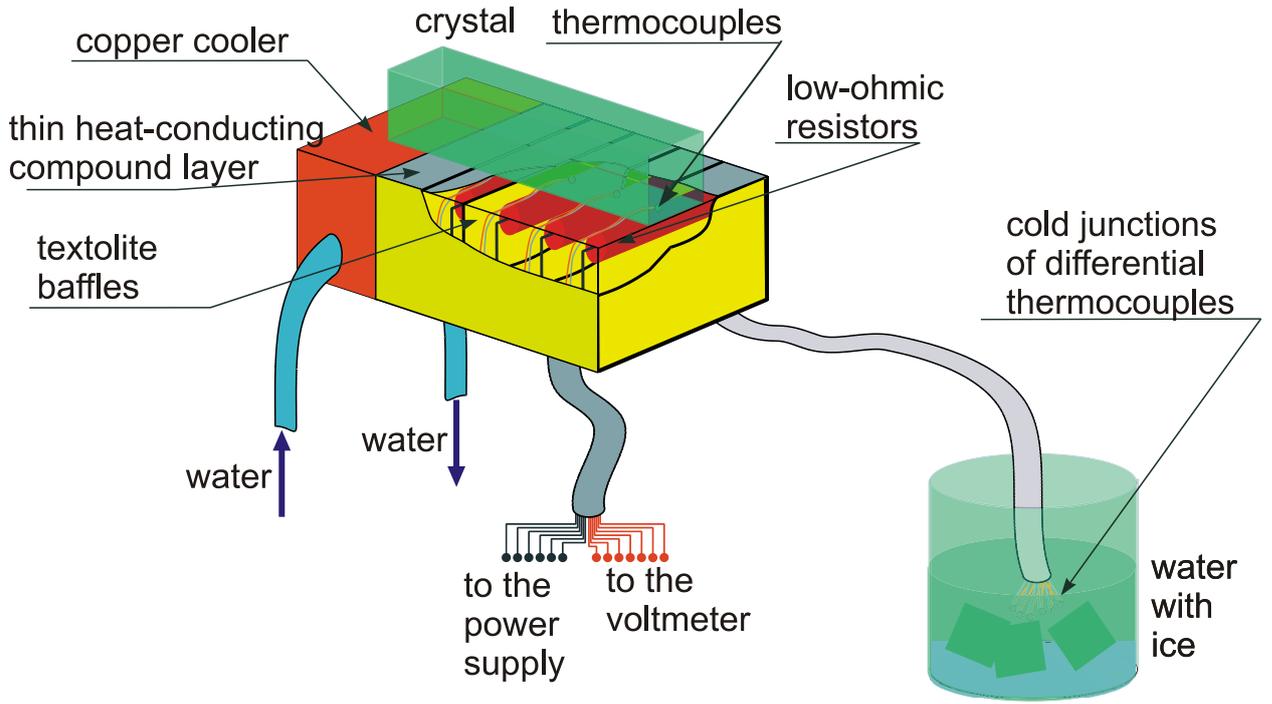

**Fig. 8. Five-section heater ensuring the inhomogeneous heating of the crystal.**
Each section contains 15-Ω resistors as heating elements and copper–constantan thermocouples. The cold junctions of thermocouples are located outside and are immersed in the mixture of water and ice in order to fix their temperature at 0˚ C. The feed circuit makes it possible to independently vary the voltage/power at each resistor.

where $\tilde{\Omega}(T)$ is the solution of the equation $\Delta k(\Omega, T) = 0$. The Sellmeier formulas [56] can be used to determine the dependence $\Delta k(\Omega, T)$.

To create a given temperature gradient along the crystal, a five-section heater was created (see Fig. 8). In order to ensure the maximum temperature gradient, a copper radiator with a one-through cooling water system was placed on one of the sides of the heater. Low-ohmic resistors were used as heating elements. To control the temperature in each section of the heater and in the radiator, thermocouples were located near the surface on which the crystal was placed. Voltages on all resistors could be controlled independently, thus ensuring the control of the temperature distribution $T(z)$ along the crystal. To ensure the thermal contact, each section was filled with a heat-conducting compound and the sections were separated by textolite plates 1.2 mm in thickness.

Note that one of the demerits of this method is the impossibility of controlling the temperature inside the crystal. Even if the pump beam maximally approaches the surface of the heater, the temperatures inside the crystal and on its surface can be strongly different at temperatures above 100°C.

Figure 9 shows the layout of the experimental setup. An Ar laser operating in the continuous regime produced a 351.1-nm pump beam with an angular divergence of about 0.2 mrad. The beam was guided to the optical system using prism P separating the necessary spectral mode and mirror M. A portion of light passed through the mirror was detected by photoresistor D1 to control the pump power. After the passage through the vertically oriented polarization prism V, the laser beam arrives at a potassium dihydrogen phosphate (KDP) crystal cut at an angle 50° to the optical axis in order to ensure the collinear degenerate matching condition. The crystal is placed on the heater and the temperature distribution $T(z)$. can be created along the crystal by applying a voltage on different sections. Filter F and horizontally oriented polarization prism H intercept parasitic pump radiation and luminescence and transmit spontaneous parametric scattering radiation. Objective O located downstream focuses radiation onto the input slit of the ISP-

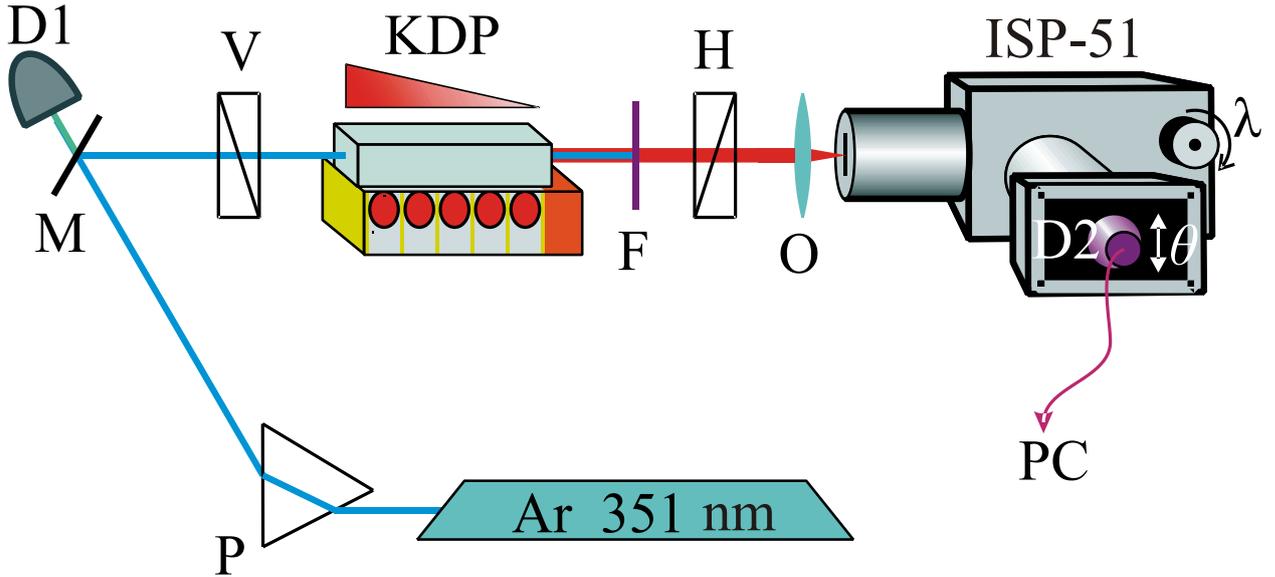

Fig. 9. Layout of the experimental setup.

51 spectrograph. Detector D2, which is a silicon avalanche photodiode operating in the photon count regime, is placed behind the spectrograph. Since the slit of the spectrograph is in the focal plane of the objective, the objective ensures the transition from the angle to the coordinate; therefore, angular scanning can be performed by moving the detector through the vertical. Wavelength scanning is performed by rotating the prism of the spectrograph.

Three series of the experiments were carried out.

In the first series, we determined the variation of the widths of the angular and frequency spectra of the biphoton field generated in the inhomogeneously heated crystal under a small change in the angle between its optical axis and the pump propagation direction. The angular spectrum was recorded in the frequency degenerate regime, the frequency spectrum was detected in the collinear regime, and the temperature distribution along the crystal remained unchanged. The results are shown in Fig. 10, where it is seen that the frequency broadening can be transformed to the angular broadening by varying the orientation of the crystal. At the maximum frequency broadening, the width of the angular spectrum is close to the width of the angular spectrum of the cold crystal and the frequency broadening disappears at the maximum broadening of the angular spectrum. Thus, these experiments show that the frequency and angular broadenings of the spontaneous parametric scattering spectra can be separated by appropriately choosing the orientation of the crystal.

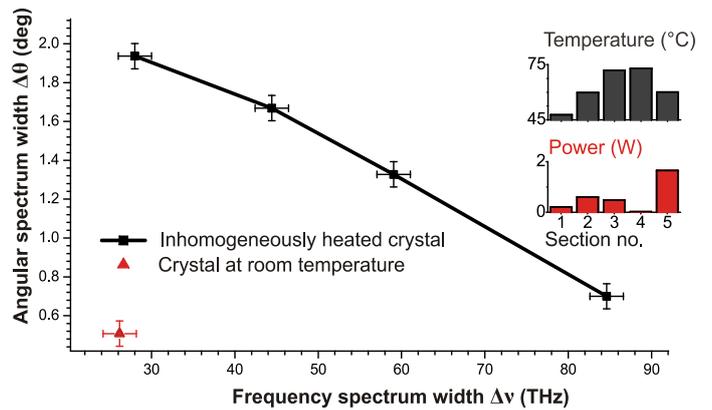

Fig. 10. Relation between the widths of the frequency, $\Delta\nu$, and angular, $\Delta\theta$, spectra of biphotons under a change in the orientation of the crystal and unchanged temperature distribution $T(z)$.

The squares are the experimental points corresponding to the inhomogeneously heated crystal. The triangle is the point corresponding to the crystal at room temperature. The histograms show the powers and temperatures of the sections of the heater (the zero section corresponds to the cooler).

In the second series, the dependence of the width of the frequency spectrum in the collinear regime on the difference between the minimum and maximum temperatures on the crystal was determined. Since the heater was longer than the crystal, its thermal contact with the radiator was not ensured; for this reason, the temperature of the radiator was disregarded when calculating this difference. The frequency spectrum was detected for two cases.

  a)  The degenerate regime of the generation of spontaneous parametric scattering. The crystal was oriented so as to ensure the maximum spectral width.

  b)  The frequency non-degenerate regime of spontaneous parametric scattering, when the difference between the frequencies nearest to the degenerate case was 84 THz.

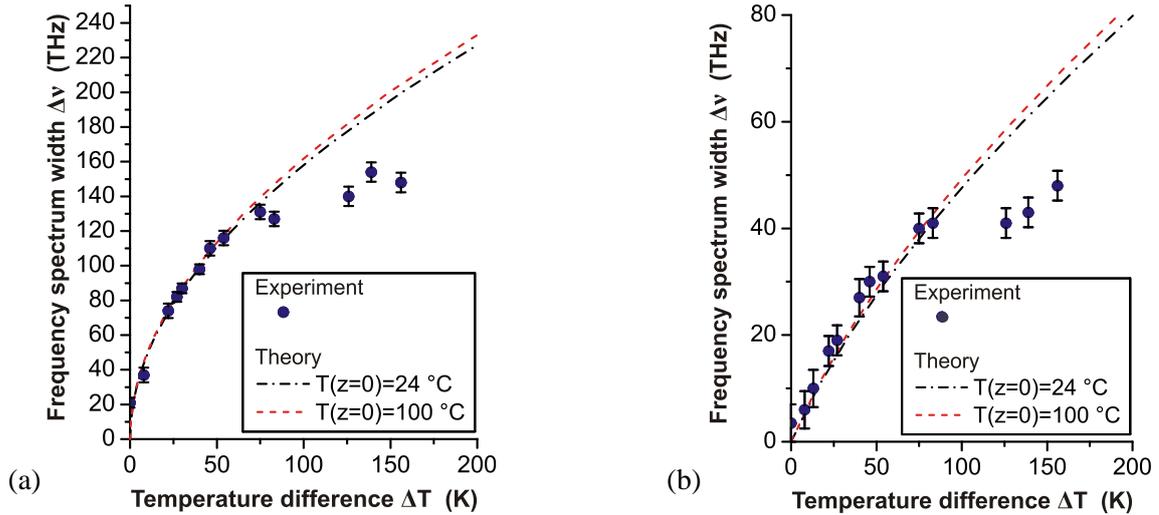

**Fig. 11. Spectral width $\Delta\nu$ versus the difference $\Delta T$ between the maximum and minimum temperatures on the crystal.**
The points are the experimental data and the dash–dotted and dashed lines are theoretical estimates by Eq. (47) for the minimum temperatures of 24 and 100°C, respectively.
(a) Degenerate case.

The results are shown in Figs. 11a and 11b, respectively. The lines are the lower theoretical estimates obtained by Eq. (47). Since the minimum temperature of the crystal increased slowly in spite of water cooling, the estimates were performed for two cases, where the minimum temperatures were 24 and 100°C. The experimental data are in good agreement with the theoretical results when the temperature gradient is no more than 75°C. Above this value, the experimental width of the spectrum is smaller than the theoretical value. The reason is that the temperature distribution inside the crystal at high temperatures (200°C and above) can be strongly different from that detected by thermocouples located in the heater near it surface. As a result of multiple heating of the crystal, the spectral width was increased from 21 to 154 THz in the degenerate regime and from 3.5 to 48 THz in the non-degenerate regime. Note that the maximum currently achieved width of the spectrum of the experimentally obtained biphoton field is 136 THz (see Fig. 7) [40].

In addition, the dependence of the integral intensity (i.e., the area under the envelope of the frequency spectrum) on the spectral width was determined for both degenerate and nondegenerate cases. In this case, the photon count rate was normalized to the pump power. The results are shown in Fig. 12. A large error in the determination of the intensity is caused by the instability of the operation of the laser and by the difficulty of the inclusion of background light from both external sources and the luminescence of the optical elements of the setup. It is seen in the plots that the integral intensity decreases. In particular, the integral intensity decreases by 84 and 72% for the cases of the maximum broadening of the spectrum up to 154 THz in the degenerate regime and up to 48 THz in the nondegenerate regime, respectively. This is apparently due to the simultaneous broadening of the angular spectrum in the transition to the significantly non-degenerate regime.

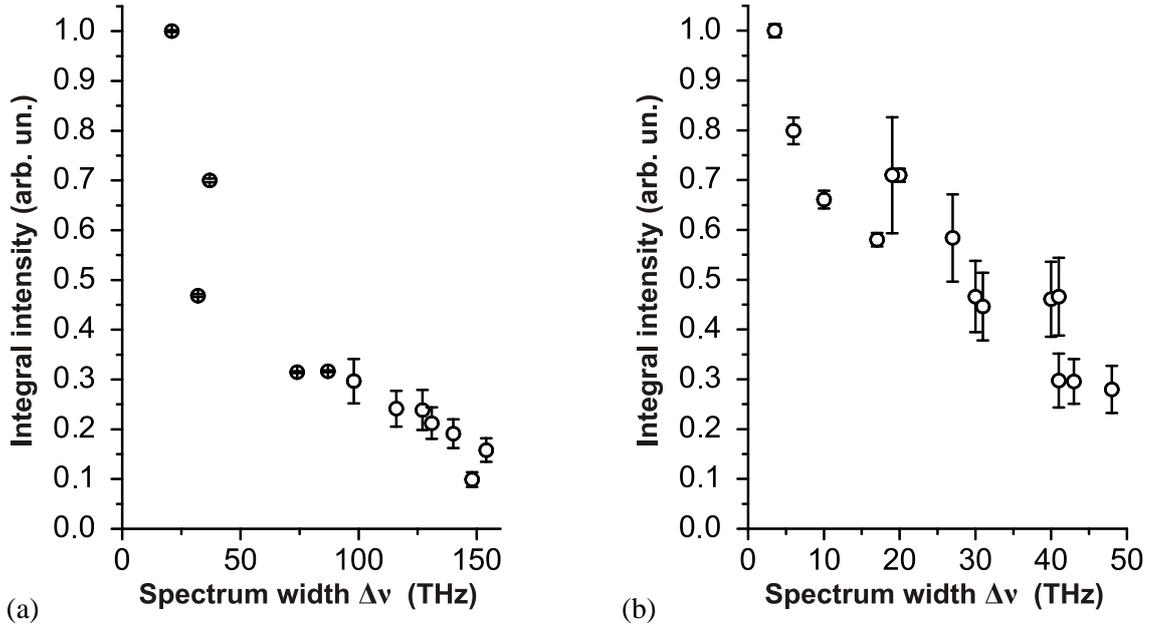

**Fig. 12. Integral intensity versus the spectral width $\Delta\nu$.**
(a) Degenerate case.
(b) Non-degenerate case.

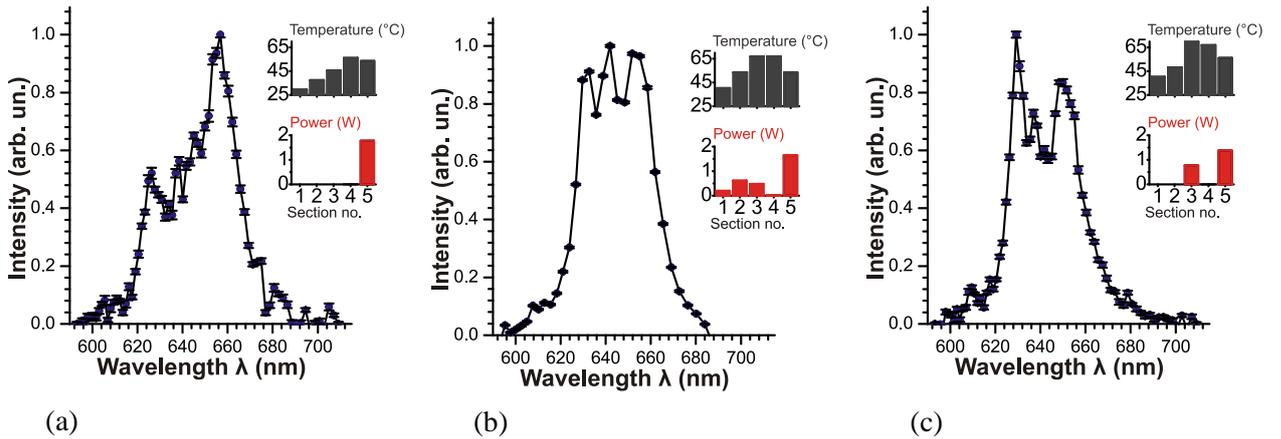

**Fig. 13. Control of the shape of the spectrum.**
The histograms show the powers and temperatures of the section of the heater (the zero section is the cooler).

In the third series of the experiments, we studied the dependence of the shape of the spectrum on the temperature distribution along the crystal. Since the shape of the spectrum near the degenerate regime depends strongly on the orientation of the crystal, the spectrum was recorded in the non-degenerate regime for the central wavelength of one of the maxima near 644 nm (at a degenerate wavelength of 702.2 nm). Figure 13 shows the three spectra obtained with various temperature distributions. In case (a), only the extreme section of the heater was switched on; for this reason, the most part of the crystal was not heated and the maximum of the spectrum is closer to the degenerate regime. In case (b), additional three sections of the heater were switched on, the crystal was heated better, and the shape of the spectrum was close to rectangular. In case (c), two sections maintain a high temperature on one half of the crystal, where the second half is not heated and the spectrum exhibits two maxima at the edges of the spectrum near and far from the degenerate regime.

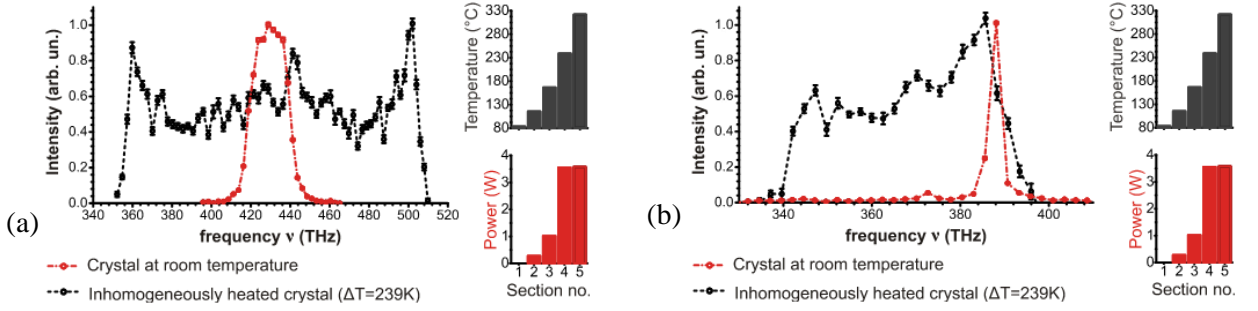

**Fig. 14. Examples of the broadening of the frequency spectrum of the biphoton field.**
(a) Degenerate case
(b) Non-degenerate case

Figures 14 and 15 exemplify the experimental frequency and angular spectra. It is worth noting that the resulting spectral widths are not maximum possible. First, a broader spectrum can be obtained with a better inhomogeneous heating (or cooling) of the crystal. Second, a better result can be obtained with other crystals. The main characteristic of the crystal necessary for such a method for broadening the spectrum is the parameter $\Delta\eta \equiv \eta(\omega_p) - \eta(\omega_s)$, introduced in Eqs. (44) and (45). It is physically determined by the temperature dependence of the dispersion of the refractive index at the frequencies $\omega_p$ and $\omega_s \approx \omega_i \approx \omega_p/2$. In the KDP crystal for the degenerate regime and $\lambda_p = 351$ nm, this parameter is $\Delta\eta = -5{,}5\times 10^{-6}\ K^{-1}$ [56]. However, a much larger broadening of the spectrum can be obtained using a medium with a larger parameter $\Delta\eta$. For example, it is an order of magnitude larger, $\Delta\eta = 2{,}4\cdot 10^{-5}\ K^{-1}$, for a lithium niobate crystal ($\lambda_p = 750$ нм, $\lambda_s \approx \lambda_i = 1500$ нм) [57].

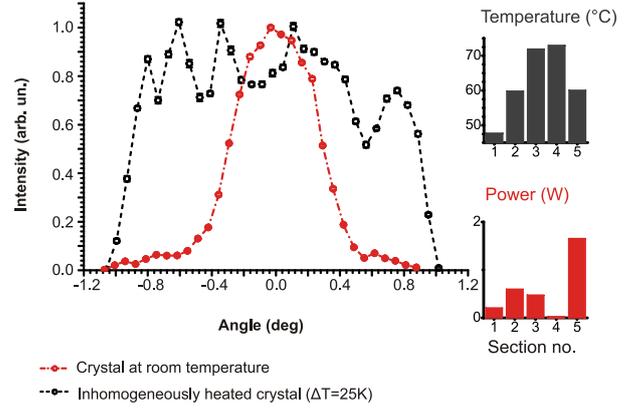

**Fig. 15. Examples of the broadening of the angular spectrum of the biphoton field.**

2. Let us consider the possibility of controlling the spectrum of the biphoton field owing to the modulation of the refractive indices induced by the electro-optical effect. Similar to the preceding case, the spectral amplitude is expressed in terms of the integral

$$F(\Omega) \propto \int_0^L dz\, exp\left[i\Delta k\left(\Omega, E(z)\right)z\right], \quad (48)$$

and the spectral width is estimated as

$$\tilde{\Omega}(E(z=0)) - \tilde{\Omega}(E(z=L)). \quad (49)$$

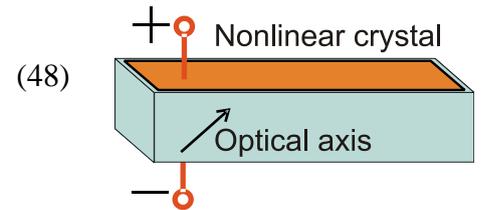

**Fig. 16. Application of the uniform electrostatic field to the KDP crystal 20 mm in length.** Electrodes are deposited on the faces of the crystal and the field inside the crystal varies depending on the voltage between the electrons.

According to Eq. (48), the method is based on the dependence $\Delta k(E)$. For this reason, to prove the possibility of controlling the spectral width, we only demonstrate a noticeable change in the spectrum when a uniform electrostatic field is applied to the crystal. To this end, electrodes were placed on the faces of the crystal (see Fig. 16) and the spectrum of spontaneous parametric scattering was detected at various voltages between the electrodes. The layout of the experimental setup shown in Fig. 9, but the crystal subjected to the uniform electrostatic field was used instead of the inhomogeneously heated crystal.

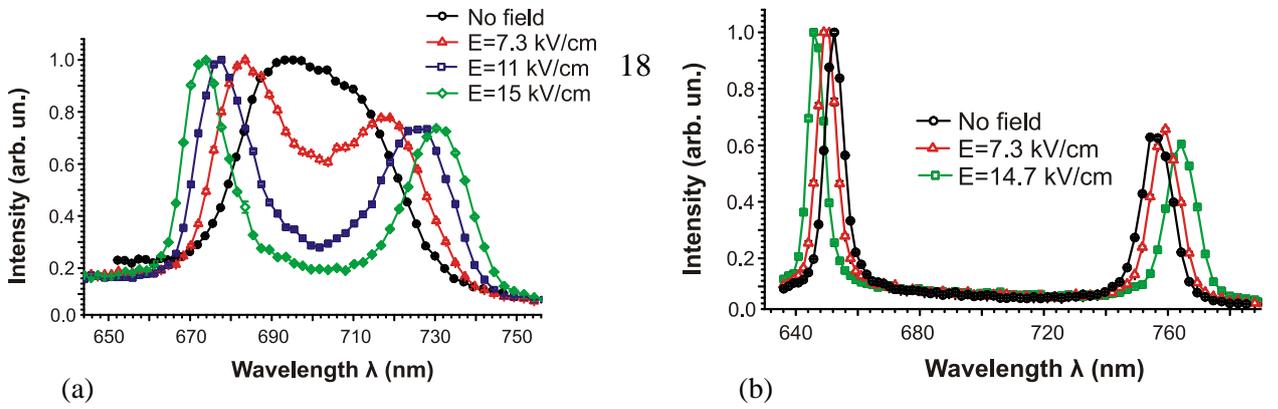

**Fig. 17.** Spectra of spontaneous parametric downconversion for various intensities of the uniform electrostatic field near the (a) degenerate and (b) non-degenerate regimes.

Figure 17 shows the spectra for various applied fields. The asymmetry of the spectra with respect to a degenerate wavelength of 702 nm is attributed to the frequency dependence of losses in the optical channel. It is seen that the applied field in the degenerate regime lifts degeneracy and leads to an increase in the distance between the peaks in the nondegenerate regime. The estimates show that the application of the spatially nonuniform static field varying along the pump propagation direction from 0 to 15 kV/cm is accompanied by an increase in the spectral width from 41 nm (25 THz) to 71 nm (43 THz) in the degenerate regime and from 6.9 to 10 THz in the nondegenerate regime. Such an inhomogeneous field distribution can be obtained by, e.g., the sectioning of electrodes.

## 6. DISCUSSION OF THE RESULTS

The method for controlling the spectrum owing to the spatial modulation of the refractive index has all advantages and disadvantages of the methods based on the use of the spatially inhomogeneous structures as compared to the methods for locally weakening the frequency dependence of the phase detuning. The spectrum of the biphoton field is not Fourier limited, complicating the use of this method to obtain fields with small second order correlation time (compression methods are necessary). At the same time, it allows a much larger broadening of spectra and is applicable even in the nondegenerate regime. Since the spectral width in this case is no longer determined by the dispersion characteristics of the crystal, there is a problem of the physical constraints for the maximum experimentally available spectral width of the biphoton field. On one hand, it is seemingly limited either by technical capabilities of varying the modulation period or by the range of the temperature drop in the field along the sample. On the other hand, as was mentioned above, the broadening of the spectrum is accompanied by a decrease in the integral intensity, which complicates its detection. This problem requires additional investigation, which is beyond the scope of this work. We also note that all known methods for controlling the frequency spectrum of the biphoton field are also applicable to control its angular spectrum; for this reason, it is reasonable to consider the problems of controlling the frequency and angular spectra together.

To conclude, we point to a number of technical advantages and disadvantages of the methods for controlling the spectrum based on the use of spatially inhomogeneous structures. The application of chirped structures is associated with a complex technological process of their manufacture. After the completion of the repolarization cycle, the spatial structure of the sample specifying the spectrum remains then unchanged. In this context, the application of the temperature gradient to the crystal seems to be a more efficient solution of the problem. In addition, not only the width, but also the shape of the spectrum can be controlled by varying the powers of the sections of the heater. At the same time, there are obvious disadvantages of this method associated with the difficulty of the determination of the real temperature inside the crystal and the creation of an arbitrary (preset) temperature distribution.

The method for broadening the spectrum owing to the nonuniform electrostatic field seems to be preferable in view of the simplicity of its implementation, although it was experimentally implemented incompletely.

## 7. CONCLUSIONS

The known methods for controlling the spectrum of the biphoton field in the presence of the spontaneous parametric downconversion of light have been analyzed. We have considered and implemented a relatively simple method for controlling the spectrum of the biphoton field by creating the longitudinal temperature gradient in the nonlinear crystal, as well as have analyzed its disadvantages and advantages as compared to other methods. The discussed method makes it possible to control the shape of the spectral distributions, as well as to reach the record broadening of the spectrum, which was 253 nm for a central wavelength of 702 nm or 154 THz, at the temperature difference between the extreme sections of the heater $\Delta T = 156 K$. The broadening of the frequency spectrum in the non-degenerate regime, as well as the broadening of the frequency and angular spectra, has been demonstrated.

## ACKNOWLEDGMENTS

In addition, the fundamental possibility of controlling the spectrum owing to the electro-optical effect has been shown.

We are deeply grateful to E.G. Yakimova and A.V. Korolev for assistance in the experiments with the electrostatic field. This work was supported by the Russian Foundation for Basic Research, project nos. 10-02-00204-a and 08-02-00741-a.